\documentclass[aps,pra,reprint,superscriptaddress]{revtex4-2}
\usepackage{graphicx}
\usepackage{dcolumn}
\usepackage{bm}
\usepackage[utf8]{inputenc}
\usepackage[T1]{fontenc}
\usepackage{booktabs, array, mathptmx, float, tabularx, booktabs, lipsum, multirow, amstext}
\usepackage{amsmath}
\usepackage{mathptmx}
\usepackage{mathrsfs}
\DeclareMathAlphabet{\mathcal}{OMS}{cmsy}{m}{n}
\usepackage{siunitx, xcolor}
\usepackage[version=4]{mhchem}

\begin{document}

\title{Feedforward Cancellation of High-Frequency Phase Noise in Frequency-Doubled Lasers}

\author{Zhen-Xing Hua}
\thanks{These authors contributed equally to this work.}
\affiliation{State Key Laboratory of Low Dimensional Quantum Physics, Department of Physics, Tsinghua University, Beijing 100084, China.}

\author{Yu-Xin Chao}
\thanks{These authors contributed equally to this work.}
\affiliation{State Key Laboratory of Low Dimensional Quantum Physics, Department of Physics, Tsinghua University, Beijing 100084, China.}

\author{Chen Jia}
\affiliation{State Key Laboratory of Low Dimensional Quantum Physics, Department of Physics, Tsinghua University, Beijing 100084, China.}

\author{Xin-Hui Liang}
\affiliation{State Key Laboratory of Low Dimensional Quantum Physics, Department of Physics, Tsinghua University, Beijing 100084, China.}

\author{Zong-Pei Yue}
\affiliation{State Key Laboratory of Low Dimensional Quantum Physics, Department of Physics, Tsinghua University, Beijing 100084, China.}

\author{Meng Khoon Tey}
\email{mengkhoon\_tey@mail.tsinghua.edu.cn}
\affiliation{State Key Laboratory of Low Dimensional Quantum Physics, Department of Physics, Tsinghua University, Beijing 100084, China.}
\affiliation{Frontier Science Center for Quantum Information, Beijing, China.}
\affiliation{Hefei National Laboratory, Hefei, Anhui 230088, China.}

\date{\today}

\begin{abstract}
The cancellation of high-frequency laser phase noise using feedforward techniques, as opposed to feedback methods, has achieved significant advancements in recent years. However, directly applying existing feedforward techniques  to laser systems based on nonlinear conversion still faces substantial challenges. Here, we propose and demonstrate a feedforward scheme that suppresses phase noise in frequency-doubled light by utilizing phase noise information of its fundamental pump. This scheme is enabled by the fact that the phase jitter of the frequency-doubled light is simply twice that of the pump, except for a first-order low-pass filtering effect introduced by the SHG enhancement cavity. Testing this method on a 420-nm frequency-doubled laser system, we realize a 25-dB suppression of the servo noise bump near 1\,MHz on the 420-nm light, and an average suppression of 30\,dB for strong injected noise ranging from 100\,kHz to 20\,MHz. This scheme shows promising potential for applications requiring blue or ultraviolet light with minimal high-frequency phase noise, such as precision control of atoms and molecules.
\end{abstract}

\maketitle

\section{Introduction}
Thanks to decades of extensive research in laser frequency stabilization and linewidth narrowing~\cite{2022-36-Hz-Integral-Linewidth-Laser-based-on-a-Photonic-Integrated-4.0-m-Coil-Resonator,2011-A-sub-40-mHz-Laser-based-on-a-Silicon-Single-Crystal-Optical-Cavity,1983-Laser-Phase-and-Frequency-Stabilization-using-an-Optical-Resonator,1971-High-Resolution-Saturation-Spectroscopy-of-the-Sodium-D-Lines-with-a-Pulsed-Tunable-Dye-Laser}, lasers with sub-kilohertz linewidth are now widely accessible. However, linewidth alone does not fully capture the spectral quality of a laser. Recent studies have highlighted the detrimental effects of high-frequency phase noise, even in lasers with remarkable linewidth. For instance, high-frequency phase noise can undermine the fidelity of quantum operations on atoms~\cite{2024-Development-of-a-High-Power-Ultraviolet-Laser-System-and-Observation-of-Fast-Coherent-Rydberg-Excitation-of-Ytterbium,2024-Benchmarking-and-Fidelity-Response-Theory-of-High-Fidelity-Rydberg-Entangling-Gates,2023-Sensitivity-of-Quantum-Gate-Fidelity-to-Laser-Phase-and-Intensity-Noise,2018-Analysis-of-Imperfections-in-the-Coherent-Optical-Excitation-of-Single-Atoms-to-Rydberg-States,2018-High-Fidelity-Control-and-Entanglement-of-Rydberg-Atom-Qubits,2014-Stochasticity-in-Narrow-Transitions-induced-by-Laser-Noise} or ions~\cite{2023-Effect-of-Fast-Noise-on-the-Fidelity-of-Trapped-Ion-Quantum-Gates,2022-Limits-on-Atomic-Qubit-Control-from-Laser-Noise,2015-Universal-Gate-Set-for-Trapped-Ion-Qubits-using-a-Narrow-Linewidth-Diode-Laser}, reduce the transfer efficiency of ultracold molecules~\cite{2024-Enhanced-Quantum-State-Transfer-via-Feedforward-Cancellation-of-Optical-Phase-Noise,2021-Efficient-Conversion-of-Closed-Channel-Dominated-Feshbach-Molecules-of-23Na40K-to-their-Absolute-Ground-State,2018-Modeling-the-Adiabatic-Creation-of-Ultracold-Polar-23Na40K-Molecules}, diminish quantum effects in optomechanical systems~\cite{2013-Laser-Noise-in-Cavity-Optomechanical-Cooling-and-Thermometry,2011-Effect-of-Phase-Noise-on-the-Generation-of-Stationary-Entanglement-Incavity-Optomechanics,2009-Phase-Noise-induced-Limitations-on-Cooling-and-Coherent-Evolution-in-Optomechanical-Systems,2008-Laser-Linewidth-Hazard-in-Optomechanical-Cooling}, and degrade the coherence of photonic-derived microwaves~\cite{2020-Coherent-Optical-Clock-Down-Conversion-for-Microwave-Frequencies-with-10−18-Instability,2018-Photonic-Microwave-Signals-with-Zeptosecond-Level-Absolute-Timing-Noise,2011-Generation-of-Ultrastable-Microwaves-via-Optical-Frequency-Division}.

Traditional frequency-stabilizing feedback techniques are ineffective at suppressing high-frequency phase noise. Due to the intrinsic time delay in feedback and its closed-loop nature, phase noise may even be amplified when the noise frequencies exceed the feedback bandwidth. Narrow-linewidth optical cavities can efficiently filter high-frequency phase noise~\cite{2024-Development-of-a-High-Power-Ultraviolet-Laser-System-and-Observation-of-Fast-Coherent-Rydberg-Excitation-of-Ytterbium,2018-High-Fidelity-Control-and-Entanglement-of-Rydberg-Atom-Qubits}. However, their applications are typically limited to transmitted optical powers below tens of milliwatts to prevent damage to the cavity mirror coatings. A promising alternative to feedback and cavity filtering is feedforward, which features a reducible time delay and an open-loop nature. This approach provides a significantly broader bandwidth than feedback, while supporting much higher optical powers than cavity filtering. Feedforward has facilitated exceptional phase noise suppression in CW lasers~\cite{2024-Pound–Drever–Hall-Feedforward-Laser-Phase-Noise-Suppression-beyond-Feedback,2024-Robust-High-Frequency-Laser-Phase-Noise-Suppression-by-Adaptive-Pound-Drever-Hall-Feedforward,2024-Measurement-and-Feed-Forward-Correction-of-the-Fast-Phase-Noise-of-Lasers,2022-Active-Cancellation-of-Servo-induced-Noise-on-Stabilized-Lasers-via-Feedforward,2019-Feedforward-Laser-Linewidth-Narrowing-Scheme-using-Acousto-Optic-Frequency-Shifter-and-Direct-Digital-Synthesizer,2012-Wide-Bandwidth-Phase-Lock-between-a-CW-Laser-and-a-Frequency-Comb-based-on-a-Feed-Forward-Configuration,2012-Wideband-Tunable-Laser-Phase-Noise-Reduction-using-Single-Sideband-Modulation-in-an-Electro-Optical-Feed-Forward-Scheme,2009-Semiconductor-Laser-Phase-Noise-Cancellation-using-an-Electrical-Feed-Forward-Scheme,1992-High-Frequency-Optical-FM-Noise-Reduction-employing-a-Fiber-Insertable-Feedforward-Technique} and frequency combs~\cite{2017-Frequency-Noise-Reduction-Performance-of-a-Feed-Forward-Heterodyne-Technique-Application-to-an-Actively-Mode-Locked-Laser-Diode,2010-Direct-Frequency-Comb-Synthesis-with-Arbitrary-Offset-and-Shot-Noise-Limited-Phase-Noise}, with records of 40\,dB suppression around 2\,MHz~\cite{2024-Pound–Drever–Hall-Feedforward-Laser-Phase-Noise-Suppression-beyond-Feedback,2024-Robust-High-Frequency-Laser-Phase-Noise-Suppression-by-Adaptive-Pound-Drever-Hall-Feedforward} and effective bandwidths above 100\,MHz~\cite{2009-Semiconductor-Laser-Phase-Noise-Cancellation-using-an-Electrical-Feed-Forward-Scheme,1992-High-Frequency-Optical-FM-Noise-Reduction-employing-a-Fiber-Insertable-Feedforward-Technique}. Notably, it has shown practical benefits in applications such as improving the state transfer efficiency of ultracold RbCs molecules~\cite{2024-Enhanced-Quantum-State-Transfer-via-Feedforward-Cancellation-of-Optical-Phase-Noise} and enhancing the control fidelity of atomic clock states~\cite{2024-Measurement-and-Feed-Forward-Correction-of-the-Fast-Phase-Noise-of-Lasers}.

Feedforward techniques have primarily been applied to red and infrared lasers. Recently, however, there has been increasing demand for highly coherent lasers in the green to ultraviolet range for applications such as quantum simulation and computation~\cite{2024-A-Dual-Species-Rydberg-Array,2020-Submicrosecond-Entangling-Gate-between-Trapped-Ions-via-Rydberg-Interaction,2020-High-Fidelity-Entanglement-and-Detection-of-Alkaline-Earth-Rydberg-Atoms,2024-Towards-Multiqudit-Quantum-Processor-based-on-a-171Yb+-Ion-String-Realizing-Basic-Quantum-Algorithms,2021-High-Fidelity-Bell-State-Preparation-with-40Ga+-Optical-Qubits}, quantum control of molecules~\cite{2021-Efficient-Conversion-of-Closed-Channel-Dominated-Feshbach-Molecules-of-23Na40K-to-their-Absolute-Ground-State,2019-Deeply-Bound-(24DJ+5S12)-Rb87-and-Rb85-Molecules-for-Eight-Spin-Couplings,2018-Experimental-Realization-of-a-Rydberg-Optical-Feshbach-Resonance-in-a-Quantum-Many-Body-System}, and optical clocks~\cite{2024-Frequency-Ratio-of-the-229mTh-Nuclear-Isomeric-Transition-and-the-87Sr-Atomic-Clock,2022-An-Optical-Atomic-Clock-based-on-a-Highly-Charged-Ion,2021-Frequency-Ratio-Measurements-at-18-digit-Accuracy-using-an-Optical-Clock-Network}. These short-wavelength light sources are typically generated via second-harmonic generation (SHG) in a nonlinear crystal pumped by high-power red or infrared light. However, implementing previous feedforward techniques directly on SHG systems presents extra challenges. On the one hand, it is arguably more advantageous to perform feedforward on the second harmonic rather than the pump, as high-power pumps in commercial systems are often inaccessible to users; even if accessible, actuators and delay fibers with demanding optical damage thresholds would be necessary. Moreover, feedforward-induced loss on the pump results in a fractional loss approximately twice as large in the SHG output due to the second-order nonlinear process. On the other hand, phase noise detection is more effective when performed on the red or infrared pump rather than on the frequency-doubled blue or ultraviolet light, as it provides a higher signal-to-noise ratio (SNR). For instance, phase noise detection using optical cavities at blue and ultraviolet wavelengths yields much worse SNR compared to red or infrared, owing to higher losses in mirror coatings and less-sensitive photo-detectors at shorter wavelengths.

In this work, we propose a feedforward scheme for suppressing high-frequency phase noise in frequency-doubled light by utilizing phase noise information of its fundamental pump. This scheme leverages the observed fact that the phase noise power spectrum of frequency-doubled light is identical to that of its pump, except for a $6\,\mathrm{dB}$ offset and a first-order low-pass filtering effect introduced by the SHG enhancement cavity. In practice, we feedforward the phase noise signal of an 840-nm pump, detected using the Pound-Drever-Hall (PDH) technique, directly to its 420-nm second harmonic through an electro-optic modulator. By calibrating the feedforward loop using an effective dual-modulation method and compensating for it with a precisely engineered loop filter, we achieve phase noise suppression in the violet laser at levels and bandwidths previously attainable only in red or infrared lasers.

\begin{figure*}[!ht]
	\centering
	\includegraphics[width=2\columnwidth]{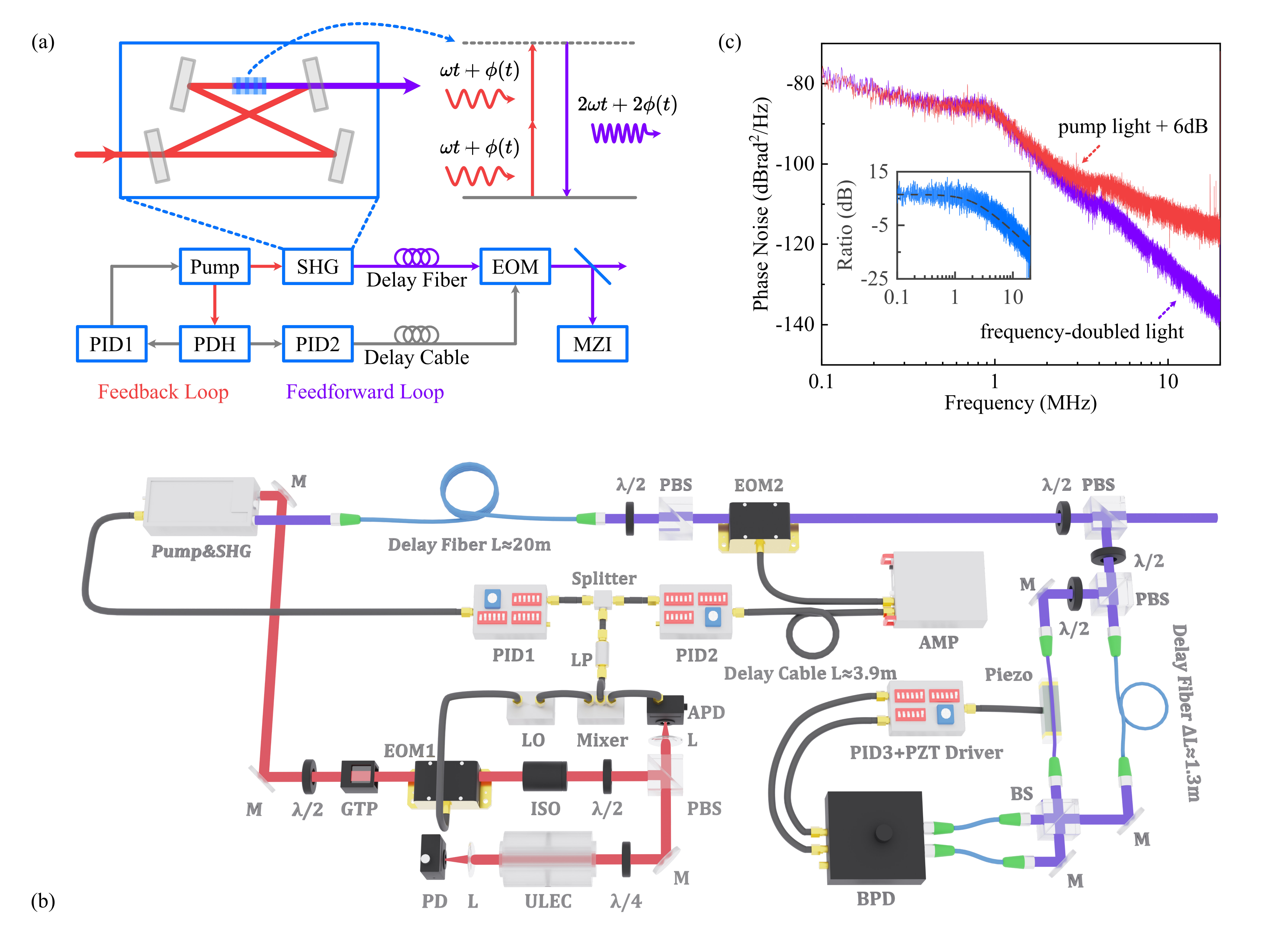}
	\caption{
(a) Main structure and (b) detailed experimental setup for suppressing high-frequency phase noise in frequency-doubled light by utilizing the phase noise information of its fundamental pump.
(c) Phase noise power spectral densities (PSDs) of the pump and the frequency-doubled light, computed from the PDH and MZI error signals, respectively. The inset shows their ratio, along with the fit to a first-order low-pass function (see text).
SHG, Second Harmonic Generation;
PDH, Pound-Drever-Hall setup;
MZI, Mach-Zehnder interferometer;
EOM, electro-optic modulator;
PID, Proportional-Integral-Differential circuit;
M, mirror;
L, lens;
$\lambda/2$, half-wave plate;
$\lambda/4$, quarter-wave plate;
PBS, polarizing beam splitter;
BS, non-polarizing beam splitter;
GTP, Glan-Taylor prism;
ISO, isolator;
ULEC, ultra-low expansion cavity;
PD, photo-detector;
APD, avalanche photo-detector;
BPD, balanced photo-detector;
LO, local oscillator;
LP, low pass filter;
AMP, high-voltage amplifier;
Piezo, stacked piezoelectric ceramics.
}
\label{setup}
\end{figure*}

\section{Phase noise relationship}
In this section, we discuss the relationship between the phase noise of the second harmonic and that of its fundamental pump. For illustration, we assume that the pump field at a specific location (e.g., $z = 0$) can be expressed as:
\begin{align}
    E_p(z=0)=\frac{1}{2} \mathcal{E} \exp \left\{\mathrm{i}\left(\omega_p t+\phi_p(t)\right)\right\}+\mathrm{c.c.},
\end{align}
where $\mathcal{E}$ and $\omega_p$ represent the complex amplitude and angular frequency of the pump, respectively, and $\phi_p(t)$ denotes the phase noise of the pump at that specific location. For simplicity, we further assume a small single-pass SHG conversion efficiency. Thus, the full expression of the pump field in a nonlinear crystal can be written as:
\begin{align}
    E_p=\frac{1}{2} \mathcal{E} \exp \left\{\mathrm{i}\left(\omega_p t-k_p z+\phi_p\left(t-k_p z / \omega_p\right)\right)\right\}+\mathrm{c.c.},
\end{align}
where $k_p$ represents the wave number of the pump light in the nonlinear crystal, and the term $k_p z / \omega_p$ accounts for the retardation of phase propagation. 

SHG arises from the nonlinear response of a medium to a strong pump field~\cite{2019-Fundamentals-of-Photonics,2008-Nonlinear-Optics} and is described by the second-order polarization $P^{(2)}$:
\begin{align}
    P^{(2)}= & \epsilon_0 \chi^{(2)} E_p^2 \nonumber \\
    = & \frac{1}{4} \epsilon_0 \chi^{(2)}\left[\mathcal{E} \mathcal{E} \exp \left\{\mathrm{i}\left(2 \omega_p t-2 k_p z+2 \phi_p\left(t-k_p z / \omega_p\right)\right)\right\}\right. \nonumber \\
    & \left.+\mathcal{E} \mathcal{E}^*+\mathrm{c.c.}\right],
\end{align}
where $\epsilon_0$ is the permittivity of vacuum, and $\chi^{(2)}$ is the second-order susceptibility of the medium. 

Locally, second-harmonic radiation is generated in synchronization with $P^{(2)}$. However, the overall SHG output is determined by the coherent sum of second-harmonic radiation generated throughout the entire crystal. In an ideal SHG system, the wave vector of the second-harmonic radiation $k_{\mathrm{SHG}}$ should match that of $P^{(2)}$, i.e., $\Delta k = 2k_p - k_\mathrm{SHG} = 0$, in order to achieve optimal SHG conversion efficiency. Obviously, in the case of perfect phase-matching, the phase noise relationship between the SHG output and the pump is:
\begin{align}\label{eq:phase_relation}
    \phi_{\mathrm{SHG}}(t)=2 \phi_p(t)
\end{align}

In the case of imperfect phase-matching, i.e., $\Delta k \neq 0$, not only will the SHG output power decrease, but Eq.~(\ref{eq:phase_relation}) will also no longer hold strictly. Nevertheless, straightforward consideration shows that Eq.~(\ref{eq:phase_relation}) remains approximately valid, provided that:
\begin{align}\label{eq:doubling_condition}
    \left|\frac{\Delta k}{2 \omega_p} \cdot L \cdot \phi_p^{\prime}(t)\right| \ll 1
\end{align}
where $L$ is the length of the nonlinear crystal and $\phi^\prime_p(t)$ represents the time derivative of $\phi_p(t)$. The physics interpretation of Eq.~(\ref{eq:doubling_condition}) is that the variation of $\phi_p(t)$ must remain much smaller than $1\,\mathrm{rad}$ over a time interval determined by the difference in transit times between the pump and second-harmonic radiation as they propagate through the nonlinear crystal. For a mismatch of $\Delta k \cdot L=2.78$, where the SHG output power is reduced to half, the corresponding time difference is on the order of $1\,\mathrm{fs}$. Thus, Eq.~(\ref{eq:phase_relation}) is expected to hold for most SHG systems.

In many SHG systems, the nonlinear crystal is typically placed in a bow-tie cavity to recycle and amplify the pump (see Fig.~\ref{setup}(a) for illustration). In addition to significantly enhancing SHG efficiency, this cavity also acts as a first-order low-pass filter on the pump. This filtering effect further ensures the fulfillment of Eq.~(\ref{eq:doubling_condition}) and thereby the validity of Eq.~(\ref{eq:phase_relation}), except that the pump's phase noise $\phi_p(t)$ should now be replaced by its low-pass-filtered counterpart $\phi_\mathrm{p,LP}(t)$. 

In short, the above discussions highlight that, apart from the low-pass filtering effect, the phase noise of frequency-doubled light is simply double in amplitude compared to that of its fundamental pump, without any changes in frequency characteristics~\cite{2019-Phase-Noise-of-Frequency-Doublers-in-Optical-Clock-Lasers,2017-Phase-Noise-Directely-Measurement-of-Optical-Second-Harmonic-Generation-in-MgOPPLN-Waveguide-based-on-the-120-degree-Phase-Difference-Interferometer,2012-A-New-Bound-on-Excess-Frequency-Noise-in-Second-Harmonic-Generation-in-PPKTP-at-the-10-19-Level}. This behavior is in agreement with the phase noise relation recognized in electronic multiplier/divider~\cite{multiplier_principle}, and has been widely applied in optical frequency comb technology~\cite{2020-Coherent-Optical-Clock-Down-Conversion-for-Microwave-Frequencies-with-10−18-Instability,2018-Photonic-Microwave-Signals-with-Zeptosecond-Level-Absolute-Timing-Noise,2011-Generation-of-Ultrastable-Microwaves-via-Optical-Frequency-Division}. However, the SHG case is more subtle, as the SHG output arises from a coherent summation process, and a low-pass filtering effect may be introduced by the SHG enhancement cavity.

\section{Experiment and Results}
The main structure of the proposed feedforward scheme is illustrated in Fig.~\ref{setup}(a). A weak portion of the pump is directed to a Pound-Drever-Hall (PDH) setup~\cite{2024-Pound–Drever–Hall-Feedforward-Laser-Phase-Noise-Suppression-beyond-Feedback,2001-An-Introduction-to-Pound-Drever-Hall-Laser-Frequency-Stabilization,1983-Laser-Phase-and-Frequency-Stabilization-using-an-Optical-Resonator} for frequency/phase noise measurement, while the remaining pump passes through a SHG system to produce frequency-doubled light. The resulting PDH error signal is used both to stabilize the pump’s frequency in a feedback loop, and to cancel the high-frequency phase noise of the frequency-doubled light in a feedforward loop. Additionally, a Mach-Zehnder interferometer (MZI)~\cite{2024-Measurement-and-Feed-Forward-Correction-of-the-Fast-Phase-Noise-of-Lasers,1992-Frequency-Domain-Analysis-of-an-Optical-FM-Discriminator} is employed to independently detect the high-frequency phase noise of the frequency-doubled light.

The detailed experimental setup is shown in Fig.~\ref{setup}(b). The red and violet lines represent the 840-nm pump and the 420-nm frequency-doubled light from a commercial laser (Pump \& SHG, Precilasers FL-SF-420-4-CW), respectively. First, the 840-nm pump is locked to an ultralow-expansion optical cavity (ULEC) with a full-width at half-maximum (FWHM) linewidth of $\Delta \nu \approx 14\,\mathrm{kHz}$, using the standard PDH technique. The 420-nm frequency-doubled light, after passing through a 20-m delay fiber and a free-space electro-optic phase modulator (EOM2, Thorlabs EO-PM-NR-C4), is partially directed into a MZI, while the remainder is available for other experiments. The MZI is quadrature-biased with a length imbalance corresponding to $\tau \approx 6.5\,\mathrm{ns}$, and the length-locking bandwidth is $f_\mathrm{lock} \approx 50\,\mathrm{kHz}$.

\subsection{Verification of Phase Noise Relationship}
\label{A}
The PDH error signal of the 840-nm pump serves as a frequency/phase discriminator at frequencies much lower/higher than $\Delta \nu$~\cite{2024-Pound–Drever–Hall-Feedforward-Laser-Phase-Noise-Suppression-beyond-Feedback,2001-An-Introduction-to-Pound-Drever-Hall-Laser-Frequency-Stabilization, 1983-Laser-Phase-and-Frequency-Stabilization-using-an-Optical-Resonator}, while the MZI error signal of the 420-nm frequency-doubled light acts as a frequency discriminator at frequencies much lower than $1/\tau \approx 133\,\mathrm{MHz}$ and much higher than $f_{\mathrm{lock}}$~\cite{2024-Measurement-and-Feed-Forward-Correction-of-the-Fast-Phase-Noise-of-Lasers,1992-Frequency-Domain-Analysis-of-an-Optical-FM-Discriminator}. As a result, the phase noise power spectral densities (PSDs) of the 840-nm pump and the 420-nm frequency-doubled light, computed from the PDH and MZI error signals, respectively, are shown in Fig.~\ref{setup}(c). Below $1\,\mathrm{MHz}$, both spectra exhibit nearly identical frequency characteristics and differ only by $6\,\mathrm{dB}$ in power. Above $1\,\mathrm{MHz}$, the first-order low-pass effect introduced by the SHG enhancement cavity becomes apparent.

This first-order low-pass effect can be verified by computing the ratio of the two power spectra and fitting the results to the function $A/(1+f^2/f_{\mathrm{SHG}}^2)$ (see inset of Fig.~\ref{setup}(c)). The fitted results yield $A \approx 4.47$, which is close to the ideal value of 4, and $f_{\mathrm{SHG}} \approx 2.09\,\mathrm{MHz}$. Notably, the $f_{\mathrm{SHG}}$ of the order of $\mathrm{MHz}$ indicates that phase noise in the $\mathrm{MHz}$ range and below will survive. However, noise in $100\,\mathrm{kHz}$ to $10\,\mathrm{MHz}$ frequency range is deleterious for many applications and cannot be effectively handled by traditional feedback techniques, thus necessitating the adoption of a feedforward technique in SHG systems.

\subsection{Optimization of Feedforward Loop}
In contrast to the feedback loop consisting of a single closed-loop path (Pump $\rightarrow$ PDH $\rightarrow$ PID1 $\rightarrow$ Pump), the feedforward loop is composed of two open-loop paths (Path 1: Pump $\rightarrow$ SHG $\rightarrow$ Delay Fiber $\rightarrow$ EOM; Path 2: Pump $\rightarrow$ PDH $\rightarrow$ PID2 $\rightarrow$ Delay Cable $\rightarrow$ EOM), as shown in Fig.~\ref{setup}(a). The open-loop nature of the feedforward loop makes it ideal for addressing high-frequency phase noise but also renders it highly sensitive to mismatches between the two paths. For instance, to achieve noise suppression greater than $40\,\mathrm{dB}$ up to $20\,\mathrm{MHz}$ using feedforward, the gain and group delay mismatch between the two paths must be less than $1\%$ and $0.08\,\mathrm{ns}$, respectively (see supplementary material for more details).

Based on the discussion in Sec.~\ref{A} and the fact that the voltage-to-phase-shift conversion ratio of EOM2 is nearly a constant, one might expect the required transfer function of feedforward Path 2 to take the form of $G/(1+\mathrm{i} f/f_\mathrm{3dB})\mathrm{e}^{-\mathrm{i} 2\pi f\tau}$, where $\mathrm{i} = \sqrt{-1}$, $G$ and $\tau$ are the overall adjustable gain and time delay, respectively, while $f_\mathrm{3dB} = f_{\mathrm{SHG}}$ is the $3\,\mathrm{dB}$ cutoff frequency of the first-order low-pass filter. However, this guess is insufficient because other components in the feedforward loop may introduce additional deviations. For example, the lack of high-power fiber modulators at short wavelengths necessitates the use of free-space modulators driven by a high-voltage amplifier (AMP), and a low-pass filter (LP) is required in the PDH setup due to its limited modulation frequency. These factors can alter $f_{\mathrm{3dB}}$ or even the form of the entire transfer function, making the matching of the feedforward loop less straightforward.

There are two possible approaches to addressing this challenge. The first is to include various filters and treat their effective cutoff frequencies, along with $G$ and $\tau$, as adjustable parameters. However, determining these parameters individually is difficult, as their influences are intertwined. The second is to test the transfer function of each component in the feedforward loop individually and then combine them. Although feasible, this approach is cumbersome, and significant discrepancies are likely to arise due to differences in power strength, impedance matching, parasitic capacitance and inductance between individual testing and combined operation.

We address this challenge by employing a dual-modulation calibration (DMC) method. The principle of this method is as follows: First, a sinusoidal phase modulation (Ch1) is applied to the pump. At the same time, an independent modulation (Ch2) of the same frequency is injected into the high-voltage amplifier (AMP) that drives EOM2, in place of the PDH signal after PID2 and the delay cable. The amplitude and phase of Ch2 are then iteratively adjusted to cancel the phase modulation in the frequency-doubled light after EOM2, thereby eliminating the observed modulation in the MZI signal. By comparing the resulting amplitude and phase of Ch2 with those of the PDH signal and repeating this process for frequencies of interest, the required transfer function of PID2 and the time mismatch between Path 1 and Path 2 can be determined (see supplemental material for more details). Finally, PID2 is set to the predetermined transfer function using a network analyzer, the length of the delay cable is adjusted based on the computed values, and the overall gain of PID2 is fine-tuned for optimal performance.

\subsection{Noise suppression performance}
\begin{figure}[!ht]
  \centering
  \includegraphics[width=0.45\textwidth]{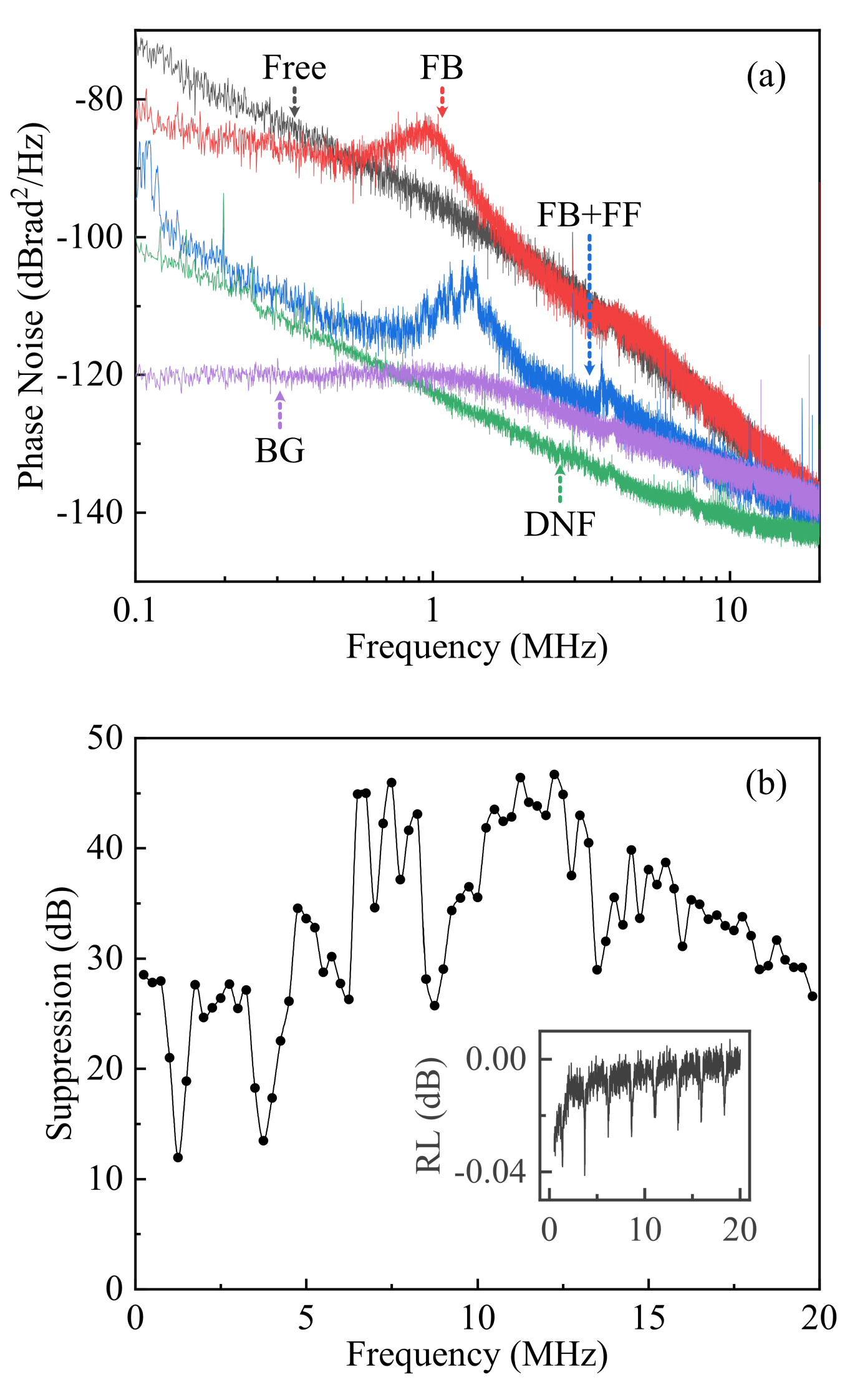}
  \caption{
(a) Phase noise power spectral densities (PSDs) of the frequency-doubled light under free-running (Free), feedback-only (FB), and combined feedback and feedforward (FB+FF) conditions. BG represents the background noise induced by feedforward Path 2, and DNF denotes the detection noise floor of the MZI.
(b) Feedforward suppression performance against strong injected sinusoidal modulations. The inset shows the return loss of the free-space EOM, whose mechanical resonance accounts for the defects in suppression performance for both intrinsic and injected noise.
}
\label{sup}
\end{figure}

Figure~\ref{sup}(a) compares the phase noise PSDs of the 420-nm frequency-doubled light under three conditions: free-running (Free), feedback-only (FB), and combined feedback and feedforward (FB+FF). The feedback process suppresses phase noise relative to the free-running case for frequencies up to $600\,\mathrm{kHz}$, but introduces a pronounced servo bump around $1\,\mathrm{MHz}$. These observations highlight the fact that a feedback loop is not only ineffective at high frequencies but also detrimental at specific frequencies. In contrast, the combined feedback and feedforward strategy outperforms both the free-running and feedback-only cases across the entire measured frequency range of $100\,\mathrm{kHz}$ to $20\,\mathrm{MHz}$. Notably, the servo bump near $1\,\mathrm{MHz}$ is attenuated by approximately $25\,\mathrm{dB}$.

The observed noise suppression when comparing the FB+FF case to the FB case is limited by several factors: above $2\,\mathrm{MHz}$, by the background noise of feedforward Path 2 (BG)~\cite{purpleline}; and below $800\,\mathrm{kHz}$, by the detection limit of the MZI. The MZI's detection limit arises from two sources: the detection noise floor of the MZI (DNF)~\cite{greenline}, and distortion induced by length-locking near $f_\mathrm{lock}$. The latter, in particular, accounts for the deviation of the FB+FF curve from the DNF curve below $100\,\mathrm{kHz}$. Importantly, the MZI's detection limit does not affect the noise suppression performance. Instead, it only limits the ability to reveal the system's actual performance, which may in fact be better than observed. In contrast, the background noise of feedforward Path 2 imposes a fundamental limit: the phase noise of the light cannot be suppressed below this level, and any noise smaller than this threshold will be amplified to match it.

To mitigate the influence of background noise and fully explore the potential of the current feedforward cancellation setup, we investigate the suppression of strong injected noise. This is achieved by applying a sinusoidal signal to the output of PID1 and measuring the resulting phase noise modulation strength of the 420-nm frequency-doubled light, both with and without feedforward. As shown in Fig.~\ref{sup}(b), the results reveal an average noise suppression above $30\,\mathrm{dB}$ across the frequency range of $100\,\mathrm{kHz}$ to $20\,\mathrm{MHz}$. At certain frequencies, the suppression even exceeds $40\,\mathrm{dB}$. The fact that feedforward cancellation remains effective up to $20\,\mathrm{MHz}$, well beyond $f_{\mathrm{SHG}}$, highlights the efficacy of the DMC method and the associated compensation strategy.

Despite the effective suppression performance for both intrinsic and injected noise over a wide frequency range, significant fluctuations occur at specific frequencies. These fluctuations arise because the voltage applied to the EOM modifies its refractive index not only through the electro-optic effect but also via mechanical vibrations induced by the piezoelectric effect~\cite{1985-Lithium-Niobate-Summary-of-Physical-Properties-and-Crystal-Structure}. The latter leads to a voltage-to-phase-shift conversion ratio that depends on mechanical resonance. For the free-space EOM used in this work, these mechanical resonances are in the MHz range and can be clearly identified by measuring the return loss of the EOM (inset of Fig.~\ref{sup}(b)). Possible solutions to this problem include reducing the size of the EOM crystal to shift the resonance frequencies higher, or altering the shape of the crystal to decrease the quality factor of the mechanical resonance.

Finally, the feedforward bandwidth in our demonstration is mainly limited by the PDH technique: at higher frequencies by the modulation frequency of the EOM, and at lower frequencies by the linewidth of the ULEC~\cite{2024-Pound–Drever–Hall-Feedforward-Laser-Phase-Noise-Suppression-beyond-Feedback,2024-Robust-High-Frequency-Laser-Phase-Noise-Suppression-by-Adaptive-Pound-Drever-Hall-Feedforward}. At higher frequencies, no significant phase noise is expected in the frequency-doubled light because of the first-order low-pass filtering. At lower frequencies, particularly in the $\mathrm{kHz}$ range and below, the delay fiber may introduce additional phase noise due to environmental perturbations. For applications requiring rigorous low-frequency phase noise control, additional fiber noise cancellation techniques should be employed~\cite{1994-Delivering-the-Same-Optical-Frequency-at-Two-Places-Accurate-Cancellation-of-Phase-Noise-introduced-by-an-Optical-Fiber-or-other-Time-Varying-Path}.

\section{Conclusions and outlooks}
We propose and demonstrate a feedforward scheme for suppressing high-frequency phase noise in frequency-doubled light, by utilizing phase noise information of its fundamental pump. This scheme is enabled by the observed fact that the phase noise power spectrum of frequency-doubled light is identical to that of its pump, except for a first-order low-pass filtering effect introduced by the SHG enhancement cavity and a $6\,\mathrm{dB}$ offset. By calibrating the feedforward loop using an effective dual-modulation method and compensating for it with a precisely engineered loop filter, we realize a 25-$\mathrm{dB}$ suppression of the servo noise bump near $1\,\mathrm{MHz}$ on the 420-nm light, and an average suppression of $30\,\mathrm{dB}$ for strong injected noise ranging from $100\,\mathrm{kHz}$ to $20\,\mathrm{MHz}$.

While this feedforward scheme is demonstrated by detecting the pump's phase noise using the PDH technique and canceling the frequency-doubled light's phase noise with an EOM, it can be readily extended to other frequency/phase noise modulators, such as acousto-optic modulator (AOM)~\cite{2019-Feedforward-Laser-Linewidth-Narrowing-Scheme-using-Acousto-Optic-Frequency-Shifter-and-Direct-Digital-Synthesizer,2012-Wide-Bandwidth-Phase-Lock-between-a-CW-Laser-and-a-Frequency-Comb-based-on-a-Feed-Forward-Configuration}, as well as to other phase noise detection methods, including self-homodyne/heterodyne detection such as MZI~\cite{2024-Measurement-and-Feed-Forward-Correction-of-the-Fast-Phase-Noise-of-Lasers,1992-Frequency-Domain-Analysis-of-an-Optical-FM-Discriminator}, and heterodyne detection with a cavity-filtered light~\cite{2022-Active-Cancellation-of-Servo-induced-Noise-on-Stabilized-Lasers-via-Feedforward,2019-Simple-Phase-Noise-Measurement-Scheme-for-Cavity-Stabilized-Laser-Systems}, reference CW laser~\cite{2017-Frequency-Noise-Reduction-Performance-of-a-Feed-Forward-Heterodyne-Technique-Application-to-an-Actively-Mode-Locked-Laser-Diode} or frequency comb~\cite{2012-Wide-Bandwidth-Phase-Lock-between-a-CW-Laser-and-a-Frequency-Comb-based-on-a-Feed-Forward-Configuration}. Furthermore, the scheme can also be adapted to other frequency conversion processes, such as third-harmonic generation and fourth-harmonic generation.

This scheme shows promising potential for a variety of applications requiring blue or ultraviolet light with minimal high-frequency phase noise. For example, it could facilitate the use of ultraviolet light for the adiabatic creation of ultracold molecules~\cite{2021-Efficient-Conversion-of-Closed-Channel-Dominated-Feshbach-Molecules-of-23Na40K-to-their-Absolute-Ground-State}, providing additional and potentially more effective transition pathways beyond red or infrared ones. With regard to the recent race using Rydberg atoms for quantum computation, our scheme may facilitate high-quality single-photon Rydberg excitation~\cite{2024-Mitigating-the-Noise-of-Residual-Electric-Fields-for-Single-Rydberg-Atoms-with-Electron-Photodesorption,2020-Transverse-Field-Ising-Dynamics-in-a-Rydberg-Dressed-Atomic-Gas,2020-High-Fidelity-Entanglement-and-Detection-of-Alkaline-Earth-Rydberg-Atoms,2016-Entangling-Atomic-Spins-with-a-Rydberg-Dressed-Spin-Flip-Blockade} as opposed to the more traditional two-photon excitation~\cite{2024-A-Dual-Species-Rydberg-Array,2020-Submicrosecond-Entangling-Gate-between-Trapped-Ions-via-Rydberg-Interaction}. The former should allow lower gate infidelities, as it is free from decoherence caused by photon scattering from the intermediate state.

\begin{acknowledgments}
This work is supported by the National Natural Science Foundation of China (NSFC) (Grant Nos. 12234012 and W2431002) and Innovation Program for Quantum Science and Technology (2021ZD0302104).
\end{acknowledgments}

\newpage

\bibliography{ms}

\begin{thebibliography}{60}
\bibitem [{pur()}]{purpleline}%
  \BibitemOpen
  \href@noop {} {}\bibinfo {note} {The background noise of feedforward Path 2 is obtained by measuring the output noise of the AMP with the feedforward loop active but the PDH feedback disabled, and then multiplying it by the voltage-to-phase-shift conversion ratio of EOM2.}\BibitemShut {Stop}%
\bibitem [{gre()}]{greenline}%
  \BibitemOpen
  \href@noop {} {}\bibinfo {note} {The detection noise floor of the MZI is obtained by sequentially blocking one of the light paths after the first beam splitter, measuring the output noise of the BPD for each case, and then summing the results.}\BibitemShut {Stop}%
\end{thebibliography}

\end{document}


\title{Supplementary Material for \\
"Feedforward Cancellation of High-Frequency Phase Noise in Frequency-Doubled Lasers"}

\maketitle

This supplemental material provides additional details on: (1) the manufacturers and models of the essential components employed in this work; (2) the tolerance of feedback and feedforward systems to gain and phase mismatches; (3) the dual-modulation calibration (DMC) method used to calibrate the feedforward loop.

\section{Manufacturers and Models}
The manufacturers and models of the essential components employed in the detailed experimental setup shown in Fig.~1(b) of the main text are as follows:
Pump \& SHG: Precilasers FL-SF-420-4-CW;
ULEC: Homemade, FWHM linewidth $\sim 14\,\mathrm{kHz}$;
EOM1: Homemade, resonant at $43.76\,\mathrm{MHz}$;
EOM2: Thorlabs EO-PM-NR-C4;
Delay fiber: Coherent PM-S405-XP;
APD: Thorlabs APD430A/M;
BPD: Thorlabs PDB450A;
PID1: Homemade;
PID2: Homemade, $0.1\,\mathrm{dB}$ bandwidth > $20\,\mathrm{MHz}$, with an optional filtering network;
PID3: Homemade;
AMP: Aigtek ATA-1200C;
LO: Homemade, oscillator at $43.76\,\mathrm{MHz}$;
Mixer: Mini-Circuits ZX05-1MHW-S+;
LP: Mini-Circuits SLP-30+;
Splitter: Mini-Circuits ZFRSC-42-S+;
Piezo: CoreMorrow Pst150/3.5*3.5/20H;

\section{Tolerance of Feedback \& Feedforward}
\begin{figure}[!htbp]
	\centering
	\includegraphics[width=1\textwidth]{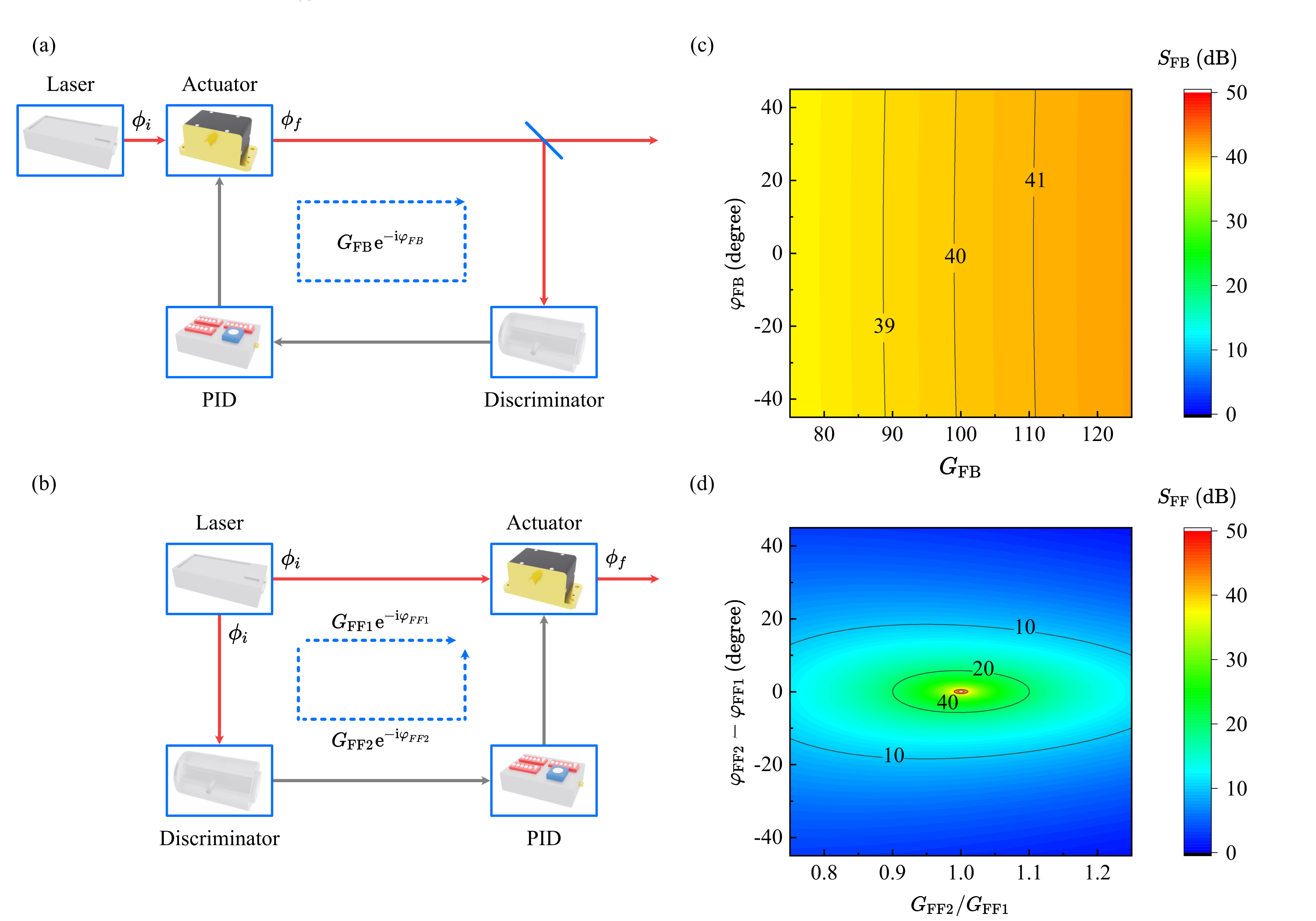}
	\caption{
(a) and (b) Typical setups for laser phase noise cancellation using feedback (FB) and feedforward (FF), respectively.
(c) and (d) Noise suppression performance of FB and FF with respect to variations in gain and phase shift, respectively.
}
	\label{tolerance}
\end{figure}

The typical setups for laser phase noise cancellation using feedback (FB) and feedforward (FF) are shown in Fig.~\ref{tolerance}(a) and (b), respectively. Although both configurations employ nearly identical components for noise detection (discriminator), processing (PID), and actuation (actuator), their characteristics differ significantly due to their distinct loop topologies. These differences include the gain and phase requirements for optimal suppression, as well as their tolerance to mismatches.

In the feedback configuration shown in Fig.~\ref{tolerance}(a), a closed-loop path is established: actuator $\rightarrow$ discriminator $\rightarrow$ PID controller $\rightarrow$ actuator. Initially, the light exhibiting phase noise $\phi_i$ is modified by the actuator, resulting in phase noise $\phi_f$. The phase noise $\phi_f$ is then detected by the discriminator, and the corresponding error signal is processed by the PID controller. The controller output is fed back to the actuator to modify the light, thereby completing the feedback loop. As a result, the relationship between the initial and final phase noise in the feedback system can be expressed as:
\begin{align}
    \phi_f=\phi_i-G_{\mathrm{FB}} \mathrm{e}^{-\mathrm{i} \varphi_{\mathrm{FB}}} \phi_f
\end{align}
where $G_{\mathrm{FB}}$ and $\varphi_{\mathrm{FB}}$ represent the total gain and phase of the feedback path. Therefore, the active noise suppression performance, with ($G_{\mathrm{FB}} \neq 0$) and without ($G_{\mathrm{FB}} = 0$) activation of the feedback loop, is given by:
\begin{align}
    S_{\mathrm{FB}}=20 \lg \left|1+G_{\mathrm{FB}} \mathrm{e}^{-\mathrm{i} \varphi_{\mathrm{FB}}}\right|
    \label{feedback}
\end{align}

In contrast, the feedforward configuration shown in Fig.~\ref{tolerance}(b) employs two open-loop paths: path 1 (laser $\rightarrow$ actuator) and path 2 (laser $\rightarrow$ discriminator $\rightarrow$ PID controller $\rightarrow$ actuator). The light with initial phase noise $\phi_i$ is split into two paths: in path 1, the light undergoes a time delay and may experience doubling or filtering (see main text); in path 2, the phase noise $\phi_i$ is detected by the discriminator, and the corresponding error signal is processed by the PID controller, as in the feedback configuration. However, the controller output is fed forward to an actuator that has not previously acted on the light, resulting in light with final phase noise $\phi_f$. As a result, the relationship between the initial and final phase noise in the feedforward system can be expressed as:
\begin{align}
    \phi_f=G_{\mathrm{FF} 1} \mathrm{e}^{-\mathrm{i} \varphi_{\mathrm{FF} 1}} \phi_i-G_{\mathrm{FF} 2} \mathrm{e}^{-\mathrm{i} \varphi_{\mathrm{FF} 2}} \phi_i
\end{align}
where $G_{\mathrm{FF}1}$, $\varphi_{\mathrm{FF}1}$, $G_{\mathrm{FF}2}$ and $\varphi_{\mathrm{FF}2}$ represent the total gain and phase of the path 1 and path 2, respectively. Therefore, the active noise suppression performance, with ($G_{\mathrm{FF}2} \neq 0$) and without ($G_{\mathrm{FF}2} = 0$) activation of the feedforward loop, is given by:
\begin{align}
    S_{\mathrm{FF}}=-20 \lg \left|1-\frac{G_{\mathrm{FF} 2} \mathrm{e}^{-\mathrm{i} \varphi_{\mathrm{FF} 2}}}{G_{\mathrm{FF} 1} \mathrm{e}^{-\mathrm{i} \varphi_{\mathrm{FF} 1}}}\right|
    \label{feedforward}
\end{align}

It is clear from Eq.~(\ref{feedback}) and Eq.~(\ref{feedforward}) that, for optimal noise suppression performance, feedback requires the highest gain ($G_{\mathrm{FB}} \gg 1$), whereas feedforward demands precise gain and phase matching ($G_{\mathrm{FF}2}/G_{\mathrm{FF}1} = 1$, $\varphi_{\mathrm{FF}2} - \varphi_{\mathrm{FF}1} = 0$). As shown in Fig.~\ref{tolerance}(c) and (d), both configurations can achieve a given suppression level, such as $40\,\mathrm{dB}$. However, their tolerance to gain and phase mismatches differs significantly. Feedback systems are relatively insensitive to small variations in gain $G_{\mathrm{FB}}$ and phase $\varphi_{\mathrm{FB}}$, whereas feedforward systems are highly sensitive to such mismatches. For example, to maintain a noise suppression level exceeding $40\,\mathrm{dB}$, even with perfect phase matching ($\varphi_{\mathrm{FF2}} - \varphi_{\mathrm{FF2}} = 0$), the gain mismatch must satisfy $|G_{\mathrm{FF2}} / G_{\mathrm{FF1}} - 1| \leq 0.01$.

This distinct tolerance to gain and phase mismatches explains why a standard commercial PID controller, with empirically tuned preset parameters, is generally sufficient for feedback systems. However, it often fails to provide the precision required for feedforward systems, especially when effective suppression is needed across a wide frequency range rather than at a single frequency. Accurately calibrating and compensating a feedforward loop presents significant challenges (see main text), highlighting the importance of the dual-modulation calibration method described below.

\section{Dual Modulation Calibration Method}
\begin{figure*}[!htbp]
	\centering
	\includegraphics[width=1\textwidth]{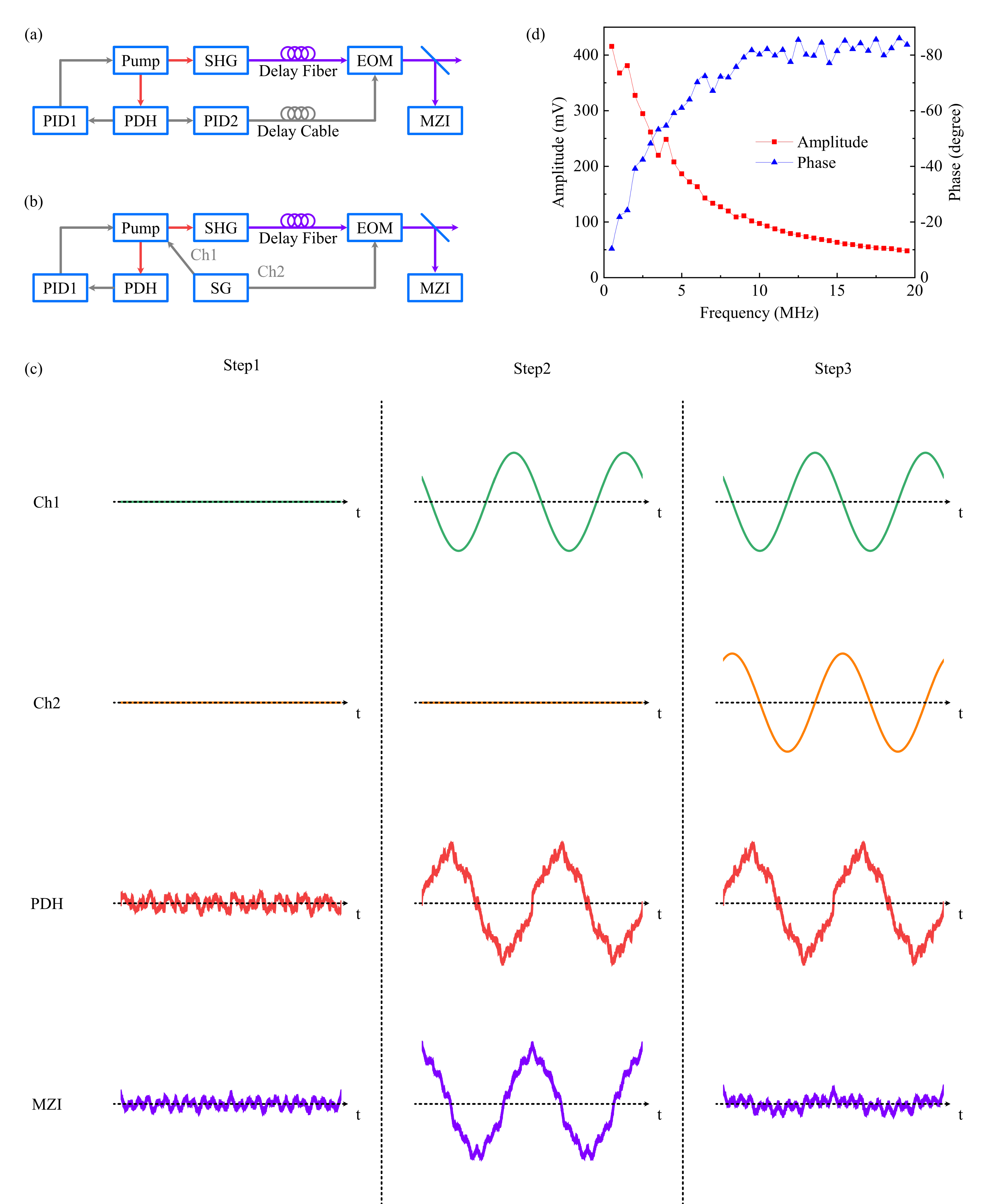}
	\caption{
(a) Schematic for feedforward cancellation of high-frequency phase noise in a frequency-doubled laser used in this work.
(b) Modifications made to (a) for calibrating the feedforward loop using the DMC method.
(c) Procedures of the DMC method and the corresponding waveforms.
(d) Amplitude and phase difference obtained using the DMC method.
}
	\label{DMC}
\end{figure*}

To address the challenges associated with matching the feedforward loop, we introduce a dual-modulation calibration (DMC) method. We begin by temporarily replacing PID2 and the delay cable with a dual-channel signal generator (SG), as shown in Fig.~\ref{DMC}(a) and (b). Following this, the DMC method is carried out in three steps, as outlined in Fig.~\ref{DMC}(c):

Step 1: Obtain the PDH signal of the 840-nm pump (serving as the in-feedforward-loop error signal) and the MZI signal of the 420-nm frequency-doubled light (serving as the out-of-feedforward-loop error signal ) without injecting any phase modulation.

Step 2: Inject a sinusoidal phase modulation at frequency $f$ into the pump via Channel 1 (Ch1) of the SG. This introduces phase modulation in both the pump and the frequency-doubled light, resulting in intensity modulation in both the PDH and MZI signals. For simplicity, we adjust the amplitude of Ch1 to ensure that the modulation strength in the PDH signal remains a constant level well above its detection noise floor, without significantly disturbing the overall PDH setup.

Step 3: Modulate the high-power amplifier that drives the EOM, using another sinusoidal signal at the same frequency $f$ from Channel 2 (Ch2) of the SG. Iteratively adjust the amplitude and phase of Ch2 to cancel the phase modulation in the frequency-doubled light after EOM2, thereby eliminating the observed modulation in the MZI signal

As a result, the optimized amplitude of Ch2 is proportional to the required gain of PID2, while the phase difference between the optimized phase of CH2 and the phase of PDH signal gives the required total phase shift, including contributions from both PID2 and the delay cable. Finally, repeat this procedure for all relevant frequencies from $100\,\mathrm{kHz}$ to $20\,\mathrm{MHz}$ to complete the calibration. 

The resulting amplitude of Ch2 and the phase difference between Ch2 and the PDH signal are shown in Fig.~\ref{DMC}(d). The results exhibit a profile consistent with a first-order low-pass filtering effect, although some defects introduced by the feedforward EOM arise. The positions of these defects correspond to those observed in the return loss measurement shown in the inset of Fig.~2(b) in the main text. We ignore these fluctuations as they are significant only at certain frequencies and are generally difficult to compensate for with standard electronic circuits. Thus, the main concerns are the low-pass filtering curve and the relative time delay, which can be obtained by fitting the amplitude and phase curves to the following transfer function:
\begin{align}
    G=A \cdot \frac{1}{1+\mathrm{i}\left(f / f_{3 \mathrm{dB}}\right)} \cdot \mathrm{e}^{-\mathrm{i} 2 \pi f \tau}
\end{align}
\begin{align}
    |G|=A \sqrt{\frac{1}{1+\left(f / f_{3\mathrm{dB}}\right)^2}}, \quad \arg (G)=-\arctan \left(\frac{f}{f_{3\mathrm{dB}}}\right)-2 \pi f \tau
\end{align}

The fitting results yield $f_{3\mathrm{dB}} = 2.4\,\mathrm{MHz}$ and $\tau = 15\,\mathrm{ns}$. Remarkably, the $f_{\mathrm{SHG}} = 2.1\,\mathrm{MHz}$ derived from the phase noise power spectral densities (PSDs) in the main text, differs from the $f_{3\mathrm{dB}} = 2.4\,\mathrm{MHz}$ obtained through the DMC method. This difference arises because $f_{3\mathrm{dB}}$ accounts for both the low-pass filtering effects of the SHG enhancement cavity and those introduced by the PDH setup’s low-pass filter and the high-voltage amplifier.

At last, while the manual repetition of the calibration steps may seem tedious, the clarity of the method makes it well-suited for automation with advanced high-speed signal processing electronics. Given its demonstrated effectiveness and potential for simplicity, we believe it will be widely applicable in the future for calibrating complex feedforward loops in actual operating conditions.

\bibliography{supplement}